\documentclass[aps,prd,a4paper,reprint,amsmath,amssymb,nofootinbib]{revtex4-2}

\usepackage[colorlinks=true, citecolor=teal]{hyperref}
\usepackage[capitalise]{cleveref}
\usepackage{scalerel}
\usepackage{mathtools}
\usepackage{enumitem}
\usepackage{bm}
\usepackage{soul}

\usepackage{parskip}

\usepackage[utf8]{inputenc}
\usepackage{fourier}
\usepackage[scale=0.8]{GoMono}
\usepackage[T1]{fontenc}
\usepackage{microtype}

\usepackage{listings}
\usepackage[most]{tcolorbox}
\usepackage{lstbayes}
\usepackage{mdframed}
\usepackage{textcomp}
\usepackage{accsupp}
\usepackage{soul}

\definecolor{white}{HTML}{FFFFFF}
\definecolor{black}{HTML}{000000}
\definecolor{blue}{HTML}{0000FF}
\definecolor{magenta}{HTML}{FF00FF}
\definecolor{violet}{HTML}{6A5ACD}
\definecolor{cyan}{HTML}{008A8C}
\definecolor{gray}{HTML}{BEBEBE}
\definecolor{green}{HTML}{2E8B57}
\definecolor{bordeaux}{HTML}{A52A2A}
\definecolor{red}{HTML}{FF0000}
\definecolor{yellow}{HTML}{FFFF00}
\definecolor{purple}{HTML}{A020F0}
\definecolor{battleshipgrey}{rgb}{0.52, 0.52, 0.51}

\newcommand{\ghost}[1]{%
\BeginAccSupp{method=escape,ActualText={}}#1\EndAccSupp{}%
}

\newcommand{\terminal}{\ghost{\textcolor{magenta}{\$}}}
\newcommand{\termout}{\color{battleshipgrey}}
\newcommand{\termskip}{\ghost{\termout\ldots\ output suppressed \ldots}}
\newcommand{\repl}{\ghost{\textcolor{magenta}{>\!>\!>}}}
\newcommand{\context}{\ghost{\textcolor{magenta}{.\!.\!.}~\textcolor{gray}{.\!.\!.\!.}}}

\setul{0.1cm}{0.15ex}
\definecolor{coderemove}{rgb}{0.82, 0.1, 0.26}
\definecolor{codeadd}{rgb}{0.3, 0.73, 0.09}
\newcommand{\codeadd}[1]{\textcolor{codeadd}{#1}}
\newcommand{\coderemove}[1]{\textcolor{coderemove}{\st{#1}}}

\newcommand{\code}[1]{\texorpdfstring{\texttt{#1}}{#1}}

\newtcolorbox{codebox}[2][]{%
    enhanced jigsaw,
    fonttitle=\color{black}\ttfamily,
    colframe=black,
    colback=white,
    colbacktitle=white,
    opacityback=0,
    opacitybacktitle=0,
    boxrule=1pt,
    top=0mm,
    arc=0mm,
    left=0mm,
    right=0mm,
    bottom=0mm,
    title={#2},#1
}

\lstdefinestyle{stanhf}{
    language=stan,
    morekeywords={std_normal,array,tuple},
}

\lstdefinestyle{bash}{
    language=bash,
    stringstyle=\color{black},
    deletekeywords={local,help,exit,for},
    literate={},
}

\lstdefinelanguage{json}{
    basicstyle=\normalfont\ttfamily\color{purple},
    literate=
     *{0}{{{\color{magenta}0}}}{1}
      {1}{{{\color{magenta}1}}}{1}
      {2}{{{\color{magenta}2}}}{1}
      {3}{{{\color{magenta}3}}}{1}
      {4}{{{\color{magenta}4}}}{1}
      {5}{{{\color{magenta}5}}}{1}
      {6}{{{\color{magenta}6}}}{1}
      {7}{{{\color{magenta}7}}}{1}
      {8}{{{\color{magenta}8}}}{1}
      {9}{{{\color{magenta}9}}}{1}
      {:}{{{\color{black}{:}}}}{1}
      {"}{{{\color{black}{"}}}}{1}
      {.}{{{\color{magenta}{.}}}}{1}
      {,}{{{\color{black}{,}}}}{1}
      {\{}{{{\color{black}{\{}}}}{1}
      {\}}{{{\color{black}{\}}}}}{1}
      {[}{{{\color{black}{[}}}}{1}
      {]}{{{\color{black}{]}}}}{1}
      {\_1}{{{\_1}}}{2}
      {\_2}{{{\_2}}}{2}
}

\lstdefinestyle{python}{%
  language     = Python,
  morekeywords = {with, as, True},
}

\newcommand{\refcite}{ref.~\cite}
\newcommand{\refcites}{refs.~\cite}

\renewcommand{\Pr}{P}
\newcommand{\pr}{p}
\newcommand{\poisson}{\operatorname{Pois}}
\newcommand{\given}{\,|\,}
\newcommand{\param}{\bm{\theta}}
\renewcommand{\vec}{\bm}
\newcommand{\pvalue}{\textit{p}-value}

\newcommand{\overclap}[3]{{\color{#3}\overbrace{#1}^{\mathclap{#2}}}}
\newcommand{\overtext}[3]{\overclap{#1}{\text{#2}}{#3}}
\newcommand{\undertext}[3]{{\color{#3}\underbrace{#1}_{\mathclap{\text{#2}}}}}
\newcommand{\breaker}[2]{
\substack{\text{#1}\\
\text{#2}}
}
\newcommand{\overtextbreak}[4]{\overclap{#1}{\breaker{#2}{#3}}{#4}}

\usepackage[british]{babel}
\usepackage[htt]{hyphenat}
\newcommand{\normfactorpg}[1]{\href{https://stan-playground.flatironinstitute.org/?project=https://gist.github.com/andrewfowlie/8734f8bf0b2841300f46ff2d4b0bba5e}{#1}}

\newcommand{\normsyspg}[1]{\href{https://stan-playground.flatironinstitute.org/?project=https://gist.github.com/andrewfowlie/323a996066a6077907626f98bde1c5cc}{#1}}

\newcommand{\gh}{https://github.com/xhep-lab/stanhf}

\begin{document}

\title{\code{stanhf}: \code{HistFactory} models in the probabilistic programming language \code{Stan}}

\author{Andrew Fowlie}
\email{andrew.fowlie@xjtlu.edu.cn}
\affiliation{\bigskip X-HEP Laboratory, Department of Physics, School of Mathematics and Physics, Xi'an Jiaotong-Liverpool University, Suzhou, 215123, China}

\begin{abstract}
\noindent In collider physics, experiments are often based on counting the numbers of events in bins of a histogram. We present a new way to build and analyze statistical models that describe these experiments, based on the probabilistic programming language \code{Stan} and the \code{HistFactory} specification. A command-line tool transpiles \code{HistFactory} models into \code{Stan} code and data files. Because \code{Stan} is an imperative language, it enables richer and more detailed modeling, as modeling choices defined in the \code{HistFactory} declarative specification can be tweaked and adapted in the \code{Stan} language. \code{Stan} was constructed with automatic differentiation, allowing modern computational algorithms for sampling and optimization.
\end{abstract}

\maketitle

\section{Introduction}

Interpreting the results of a physics experiment requires a statistical model. This model describes the relationship between a theory and potential observations, including statistical and systematic errors that could affect the outcome. In collider physics, experiments are often based on counting the numbers of events in bins of a histogram. We present a new way to build and analyze these statistical models, based on the probabilistic programming language (PPL) \code{Stan}~\cite{Carpenter2017,stan}. These computer representations of a statistical model preserve and communicate an experimental analysis and enable it to be modified or recasted. 

More generally, PPLs are languages that are designed for statistical modeling (see e.g., refs.~\cite{BMCP2021,2021NatRP...3..305B}). Like most languages, they support random number generation and their outputs can be probabilistic; significantly, however, they are designed to make it easy to express statistical models and perform statistical inference. Besides \code{Stan}, examples of PPLs include \code{WinBUGS}~\cite{Lunn2000}, \code{Turing.jl}~\cite{10.1145/3711897}, \code{PyMC}~\cite{pymc}, \code{Edward}~\cite{tran2016edward,tran2018simple}, \code{pyro/numpyro}~\cite{2018arXiv181009538B,phan2019composable}, \code{WinBUGS}~\cite{Lunn2000} and \code{RooFit/RooStats}~\cite{Antcheva:2009zz,Verkerke:2003ir,Moneta:2010pm}. In the context of high-energy physics, PPLs were recently explored in refs.~\cite{Cranmer:2021gdt,2019arXiv190703382G,Qin:2024zou,NEURIPS2019_6d19c113,elvira2022future}.

To understand the form of statistical models for these counting experiments, we start from the Poisson likelihood for data binned in a histogram or several histograms
\begin{equation}\label{eq:simple}
\Pr(\vec{o}) = \prod_i \poisson(o_i \given \lambda_i).
\end{equation}
where the index $i$ runs over all histogram bins, and $o_i$ and $\lambda_i$  represent the observed and expected numbers of events in bin $i$, respectively. There are several contributions to the expected number of events in a bin from specific backgrounds and from potential signals, such that we write
\begin{equation}
\lambda_i = \sum_j \lambda_{i\!\!j}.
\end{equation}
The nominal rates for each contribution to each bin, $\lambda_{i\!\!j}$, can be affected by systematic errors. These errors are modeled by introducing nuisance parameters $\param$. These parameters control multiplicative factors and additive terms that modify the nominal rates,
\begin{equation}
\lambda_{i\!\!j} \to  \mu_{i\!\!j} (\param) \left(\lambda_{i\!\!j} + \delta_{i\!\!j}(\param)\right)
\end{equation}
where the factor $\mu_{i\!\!j}$ represents the product of all multiplicative systematic effects  and the term $\delta_{i\!\!j}$ represents the sum of all additive systematic effects for the nominal rate for bin $i$ contribution $j$. Thus, they can be broken down into individual systematic effects,
\begin{equation}
    \mu_{i\!\!j}(\param) = \prod_k \mu_{i\!\!j\!k}(\param)\quad\text{and}\quad \delta_{i\!\!j}(\param) = \sum_k \delta_{i\!\!j\!k}(\param)
\end{equation}
where $k$ indexes individual systematic effects, and $\mu_{i\!\!j\!k}$ and $\delta_{i\!\!j\!k}$ represent individual multiplicative and systematic effects for bin $i$, contribution $j$ and effect $k$. Lastly, the parameters $\param$ might be constrained by prior knowledge or auxiliary experiments. Thus, combining these changes to \cref{eq:simple}, we arrive at
\begin{equation}
\begin{split}
    \pr(\vec{o}, \param \given K) ={}& 
\overtext{\prod_i}{Bins}{purple}
\poisson\bigl(o_i \,\big|\!
    \undertext{\sum_j}{Samples}{green} \!
    \overtextbreak{\mu_{i\!\!j}(\param)}{Multiplicative}{modifier}{blue} 
    \,\bigl[\!
    \undertext{\lambda_{i\!\!j}}{Nominal rate}{violet} \!+ 
    \overtextbreak{\delta_{i\!\!j}(\param)}{Additive}{modifier}{red}
    \bigr]
    \bigr) \\[-4.5mm]
{}&\cdot \overtext{\pr(K \given \param)}{Auxiliary}{magenta}
\end{split}\label{eq:model}
\end{equation}
where $K$ represents prior knowledge or data from auxiliary measurements.

On the other hand, rather than binning events into histograms, one could model the events using a probability density function (pdf). Suppose that we observed $o$ events with a discriminating variable, e.g.~invariant mass, $\{x_1, \ldots,x_o\}$. Rather than binning with respect to the discriminating variable as in \cref{eq:simple}, we write
\begin{equation}\label{eq:unbinned}
\pr(\{x_1, \ldots,x_o\}) =  \poisson(o \given \lambda) \prod_{i=1}^o f(x_i)
\end{equation}
where the pdf $f(x)$ models the discriminating variable for each event. Each observed event is marked by its value of the discriminating variable, and thus \cref{eq:unbinned} is known as a marked Poisson model.

Previously, the \code{HistFactory} specification was developed to represent binned models~\cite{Cranmer:1456844}. This is a declarative specification of a statistical model of the form \cref{eq:model} --- the modifiers and auxiliary term in \cref{eq:model} can be built from a set of predeclared choices. The specification was written in \code{xml} language and was intended to be interpreted by the \code{RooFit/RooStats} framework based on \code{ROOT} histogram functionality~\cite{Antcheva:2009zz,Verkerke:2003ir,Moneta:2010pm}. More recently, a \code{json} schema was designed to express \code{HistFactory} models in \code{json}~\cite{Feickert:2712783,ATLAS:2019oik}.

At the same time, the \code{Python} package \code{pyhf}~\cite{Feickert:2022lzh,pyhf_joss,pyhf} was developed to interpret and analyze \code{HistFactory} models from \code{json} files. As \code{Python} is somewhat ubiquitous compared to \code{ROOT}, this increased the portability and potentially increases the longevity of these statistical models. \code{pyhf}, furthermore, takes advantage of automatic differentiation, so that derivatives are available for optimization. The results of searches for new particles at the Large Hadron Collider (LHC)  are often preserved as \code{HistFactory} models and hosted publicly on the \code{HEPData} webpage~\cite{Maguire2017}. At present, \code{HistFactory} models and \code{pyhf} are used by the ATLAS, MicroBooNE~\cite{MicroBooNE:2023gmv}, Belle and Belle-II~\cite{Belle-II:2023esi}, and MODE~\cite{MODE:2022znx} collaborations. Bayesian analyses are possible through an interface to \code{PyMC}~\cite{pymc} in \code{bayesian\_pyhf}~\cite{Feickert:2023hhr}. 

We present a complementary tool to the successful \code{pyhf} and \code{HistFactory} based on \code{Stan}~\cite{Carpenter2017,stan}. \code{Stan} is an imperative probabilistic programming language with automatic differentiation~\cite{Gorinova_2019}, enabling models to be analyzed in a Bayesian framework using Hamiltonian Monte Carlo~(HMC;~see e.g.~\refcites{Neal:2011mrf,Betancourt:2017ebh}). This is an efficient algorithm for obtaining a sample-representation of the posterior distribution. \code{Stan} started development in 2011, and is used in industry~\cite{prophet,Taylor2017} and academia, including cognitive, social and political sciences, medical statistics, and astrophysics~\cite{Rubin:2023ovl,KAGRA:2021duu,Farr:2019twy}. As we build on \code{HistFactory}, we focus on binned models, though unbinned models such as \cref{eq:unbinned} could be programmed in \code{Stan}. 

Because \code{Stan} is an imperative language, it is more flexible and expressive than declarative specifications. This enables richer and more detailed modeling, as the modeling choices defined in the \code{HistFactory} declarative specification can be tweaked and adapted in the \code{Stan} language. This is a major difference from \code{pyhf} and \code{bayesian\_pyhf}. Although \code{Stan} was designed primarily for Bayesian modeling and computation, frequentist statistics, such as confidence intervals and \pvalue{}s, are available in \code{stanhf} through an interface with \code{pyhf}. \code{Stan} aims to offer state-of-the-art performance and supports computation on GPUs~\cite{2019arXiv190701063C}. Performance comparisons to \code{JAX}~\cite{jax2018github}, a possible computational backend in \code{pyhf}, indicate that \code{Stan} outperforms \code{JAX} on the CPU but that \code{JAX} outperforms \code{Stan} on the GPU~\cite{stancon2024}.

\section{Quickstart}\label{sec:quickstart}

The package and dependencies can be installed by
\begin{lstlisting}[style=bash]
|\terminal| pip install stanhf
\end{lstlisting}
The source code can be obtained by
\begin{lstlisting}[style=bash]
|\terminal| git clone |\gh|
\end{lstlisting}
You can obtain a \code{HistFactory} model in \code{json} specification from \code{HEPData}. E.g., using \code{curl},
\begin{lstlisting}[style=bash]
|\terminal| curl -OJLH "Accept: application/x-tar" https://doi.org/10.17182/hepdata.89408.v3/r2
|\terminal| tar -xzf HEPData_workspaces.tar.gz
\end{lstlisting}
or just browsing the landing page for the resource e.g.~\href{https://www.hepdata.net/record/resource/1935437?landing_page=true}{here}. This downloads and extracts files, including
\begin{lstlisting}[style=bash]
RegionA/BkgOnly.json  # hf model
RegionA/patchset.json  # patches to modify hf model
README.md  # metadata about model
\end{lstlisting}
for \code{HistFactory} models for searches for squarks by ATLAS at the LHC in \refcite{ATLAS:2019gdh}.

The \code{HistFactory} model \code{RegionA/BkgOnly.json} can be transpiled into \code{Stan} using the command line interface (CLI),
\begin{lstlisting}[style=bash]
|\terminal| stanhf RegionA/BkgOnly.json
\end{lstlisting}
This writes a \code{HistFactory} model as a \code{Stan} program, compiles it, and validates it. If necessary, \code{cmdstan}~\cite{cmdstan} is automatically downloaded and installed so that \code{Stan} models can be compiled. These steps, as well as metadata about the model, are written to \code{stdout}:
\begin{lstlisting}[style=bash]
hf file 'RegionA/BkgOnly.json' with no patch applied:
- 57 free parameters, 0 fixed parameters and 8 ignored null parameters
- 3 channels with 24 samples
- 839 modifiers and 431 ignored null modifiers
- Stan installed at .cmdstan/cmdstan-2.3.5
- Stan files created at RegionA/BkgOnly.stan, RegionA/BkgOnly_data.json and RegionA/BkgOnly_init.json
- Validated parameter names
- Build settings controlled at .cmdstan/cmdstan-2.3.5/build/local
- Stan executable created at RegionA/BkgOnly
- Try e.g., RegionA/BkgOnly sample num_chains=4 data file=RegionA/BkgOnly_data.json init=RegionA/BkgOnly_init.json
- Validated target
\end{lstlisting}
As indicated, three files are written to disk, and the \code{Stan} model is compiled into an executable: 
\begin{lstlisting}[style=bash]
RegionA/BkgOnly.json  # original hf model
RegionA/BkgOnly.stan  # the Stan model
RegionA/BkgOnly_data.json  # the associated data
RegionA/BkgOnly_init.json  # initial values for parameters 
RegionA/BkgOnly  # compiled executable
\end{lstlisting}
The process is summarized for simplified example \code{HistFactory} models in \cref{fig:example_normfactor,fig:example_normsys}. The simplified models include a normalization factor (denoted by \code{normfactor} in the \code{HistFactory} specification) and a normalization systematic uncertainty (denoted by \code{normsys} in the \code{HistFactory} specification), respectively. They  can be found in the \code{stanhf} source code at  \code{examples/normfactor.json} and \code{examples/normsys.json}, respectively.

\begin{figure*}
\begin{tcolorbox}[blankest, watermark text=\code{stanhf}\\\vspace{-4.5mm}\\$\scaleobj{1.5}{\Longrightarrow}$,
  watermark zoom=0.55,
  watermark color=violet,
  watermark opacity=0.25,
  clip watermark]
\begin{tcbraster}[raster columns=2,raster valign=top]
\begin{codebox}{HistFactory --- normfactor.json \vphantom{Stan --- normfactor.stan}}
\begin{lstlisting}[language=JSON,frame=none]
{
    "channels": [
        { "name": "channel",
          "samples": [
            { "name": "background",
              "data": [50.0, 60.0],
              "modifiers": []
            },
            { "name": "signal",
              "data": [5.0, 10.0],
              "modifiers": [
                { "name": "mu", "type": "normfactor", "data": null}
              ]
            }
          ]
        }
    ],
    "observations": [
        { "name": "channel", "data": [44.0, 62.0] }
    ],
    "measurements": [
        { "name": "Measurement", "config":
          {"poi": "mu", "parameters": [ 
            { "name":"mu", "bounds": [[0.0, 10.0]], "inits": [1.0] }
          ]
          }
        }
    ],
    "version": "1.0.0"
}
\end{lstlisting}%
\end{codebox}%
\begin{tcolorbox}[blankest]
\begin{tcbraster}[raster columns=1]
\begin{codebox}{Stan --- normfactor.stan \vphantom{HistFactory --- normfactor.json}}%
\begin{lstlisting}[style=stanhf,frame=none]
data {
  vector[2] nominal_channel_background;
  vector[2] nominal_channel_signal;
  int<lower=0, upper=1> fix_mu;
  real fixed_mu;
  tuple(real, real) lu_mu;
  array[2] int observed_channel;
}
parameters {
  array[1 - fix_mu] real<lower=lu_mu.1, upper=lu_mu.2> free_mu;
}
transformed parameters {
  vector[2] expected_channel_background = nominal_channel_background;
  vector[2] expected_channel_signal = nominal_channel_signal;
  real mu = fix_mu ? fixed_mu : free_mu[1];
  expected_channel_signal *= mu;
  vector[2] expected_channel = expected_channel_background + expected_channel_signal;
}
model {
  observed_channel ~ poisson(expected_channel);
}
generated quantities {
  array[2] int rv_expected_channel = poisson_rng(expected_channel);
}

||
\end{lstlisting}
\end{codebox}
\begin{codebox}{Stan data --- normfactor\_data.json}
\begin{lstlisting}[language=JSON,frame=none]
{
    "fix_mu": 0,
    "fixed_mu": 1.0,
    "lu_mu": {
        "1": 0.0,
        "2": 10.0
    },
    "nominal_channel_background": [
        50.0,
        60.0
    ],
    "nominal_channel_signal": [
        5.0,
        10.0
    ],
    "observed_channel": [
        44,
        62
    ]
}
\end{lstlisting}
\end{codebox}
\begin{codebox}{Stan initial values --- normfactor\_init.json}
\begin{lstlisting}[language=JSON,frame=none]
{
    "free_mu": [
        1.0
    ]
}
\end{lstlisting}
\end{codebox}
\end{tcbraster}
\end{tcolorbox}
\end{tcbraster}
\end{tcolorbox}
\caption{\code{stanhf} reads the declarative specification of a \code{HistFactory} model from a \code{json} file and writes a \code{Stan} model and associated data files. For simplicity, we omit the \code{functions} block of the \code{Stan} model and stripped comments and metadata from \code{stanhf} outputs. This model can be ran in your browser \normfactorpg{here} and found in the \code{stanhf} source code in \code{examples/normfactor.json}.}
\label{fig:example_normfactor}
\end{figure*}

\begin{figure*}
\begin{tcolorbox}[blankest, watermark text=\code{stanhf}\\\vspace{-4.5mm}\\$\scaleobj{1.5}{\Longrightarrow}$,
  watermark zoom=0.55,
  watermark color=violet,
  watermark opacity=0.25,
  clip watermark]
\begin{tcbraster}[raster columns=2,raster valign=top]
\begin{codebox}{HistFactory --- normsys.json \vphantom{Stan --- normsys.stan}}
\begin{lstlisting}[language=JSON,frame=none]
{
    "channels": [
        { "name": "channel",
          "samples": [
            { "name": "background",
              "data": [50.0, 60.0],
              "modifiers": []
            },
            { "name": "signal",
              "data": [5.0, 10.0],
              "modifiers": [
                { "name": "mu", "type": "normsys", "data": {"hi": 1.1, "lo": 0.9}}
              ]
            }
          ]
        }
    ],
    "observations": [
        { "name": "channel", "data": [44.0, 62.0] }
    ],
    "measurements": [
        { "name": "Measurement", "config":
          {"poi": "mu", "parameters": [ 
            { "name":"mu", "bounds": [[0.0, 10.0]], "inits": [1.0] }
          ]
          }
        }
    ],
    "version": "1.0.0"
}
\end{lstlisting}%
\end{codebox}%
\begin{tcolorbox}[blankest]
\begin{tcbraster}[raster columns=1]
\begin{codebox}{Stan --- normsys.stan \vphantom{HistFactory --- normsys.json}}%
\begin{lstlisting}[style=stanhf,frame=none]
data {
  vector[2] nominal_channel_background;
  vector[2] nominal_channel_signal;
  int<lower=0, upper=1> fix_mu;
  real fixed_mu;
  tuple(real, real) lu_mu;
  tuple(real, real) lu_channel_signal_normsys_mu;
  array[2] int observed_channel;
}
parameters {
  array[1 - fix_mu] real<lower=lu_mu.1, upper=lu_mu.2> free_mu;
}
transformed parameters {
  vector[2] expected_channel_background = nominal_channel_background;
  vector[2] expected_channel_signal = nominal_channel_signal;
  real mu = fix_mu ? fixed_mu : free_mu[1];
  expected_channel_signal *= factor_interp(mu, lu_channel_signal_normsys_mu);
  vector[2] expected_channel = expected_channel_background + expected_channel_signal; 
}
model {
  observed_channel ~ poisson(expected_channel);
  mu ~ std_normal();
}
\end{lstlisting}
\end{codebox}
\begin{codebox}{Stan data --- normsys\_data.json}
\begin{lstlisting}[language=JSON,frame=none]
{
    "fix_mu": 0,
    "fixed_mu": 1.0,
    "lu_channel_signal_normsys_mu": {
        "1": 0.9,
        "2": 1.1
    },
    "lu_mu": {
        "1": 0.0,
        "2": 10.0
    },
    "nominal_channel_background": [
        50.0,
        60.0
    ],
    "nominal_channel_signal": [
        5.0,
        10.0
    ],
    "observed_channel": [
        44,
        62
    ]
}
\end{lstlisting}
\end{codebox}
\begin{codebox}{Stan initial values --- normsys\_init.json}
\begin{lstlisting}[language=JSON,frame=none]
{
    "free_mu": [
        1.0
    ]
}
\end{lstlisting}
\end{codebox}
\end{tcbraster}
\end{tcolorbox}
\end{tcbraster}
\end{tcolorbox}
\caption{Similar to \cref{fig:example_normfactor}, though for a model with a systematic uncertainty in the normalization of a signal. For simplicity, we omit the \code{functions} and \code{generated quantities} blocks of the \code{Stan} model and stripped comments and metadata from \code{stanhf} outputs. This model can be ran in your browser \normsyspg{here} and found in the \code{stanhf} source code in \code{examples/normsys.json}.} 
\label{fig:example_normsys}
\end{figure*}

As suggested in the output, you can run this model by e.g.
\begin{lstlisting}[language=bash]
RegionA/BkgOnly sample num_chains=4 data file=RegionA/BkgOnly_data.json init=RegionA/BkgOnly_init.json
\end{lstlisting}
For further information on running an analysis, see \cref{sec:bayes,sec:freq}.

\section{Further details}

\subsection{Structure of \code{Stan} model}

\code{Stan} is a strongly typed and imperative probabilistic programming language. \code{Stan} programs, including those created by \code{stanhf}, define a statistical model through special blocks. We present example \code{Stan} programs produced from the simplified examples \code{examples/normfactor.json} and \code{examples/normsys.json} in \cref{fig:example_normfactor,fig:example_normsys}, respectively.

First, the \code{functions} block. In \code{stanhf} we use this block to define functions for interpolation and constraint terms that are required to implement the \code{HistFactory} specification.
\begin{lstlisting}[language=stan]
functions {
  // functions for interpolation and constraints
}
\end{lstlisting}
We declare the data in the \code{data} block; here, data includes experimental data, as well as nominal event numbers, and hyperparameters that define modifiers and constraint terms.  We do not define data here; it is written to a separate \code{json} file.  In the context of high-energy physics, the natural separation between a model and observed data in \code{Stan} makes data-blinding straight-forward~\cite{Chekanov:2021kwl}, as observed data can be substituted without altering the model or model file.
\begin{lstlisting}[language=stan]
data {
    // declare observed data, nominal rates etc
    // data defined in separate json file
}
\end{lstlisting}
Occasionally, we need to make transformations of the data before we begin sampling, e.g., to aggregate uncertainties for a constraint term.
\begin{lstlisting}[language=stan]
transformed data {
    // e.g. aggregate uncertainties declared in data block
}
\end{lstlisting}
Unknown parameters, associated with modifiers, are declared in the \code{parameters} block. These parameters are marginalized or profiled in an analysis.
\begin{lstlisting}[language=stan]
parameters {
    // declare unknown parameters to be optimized/marginalized
}
\end{lstlisting}
The expected event rates are computed by applying modifiers to the nominal rates in the \code{transformed parameters} block.
\begin{lstlisting}[language=stan]
transformed parameters {
    // compute e.g. expected event rates using parameters
}
\end{lstlisting}
The Poisson likelihood and constraint terms for background knowledge or auxiliary measurements are entered in the \code{model} block.
\begin{lstlisting}[language=stan]
model {
    // add Poisson distributions, constraint terms etc
}
\end{lstlisting}
Finally, we simulate data from the model in the \code{generated quantities} block. This simulated data can be used to compute posterior predictive distributions. 
\begin{lstlisting}[language=stan]
generated quantities {
    // simulate data from model
}
\end{lstlisting}

\subsection{Choices of prior}

\code{Stan} is agnostic about whether auxiliary terms in \cref{eq:model} are part of the prior or likelihood, since it is only their product that matters. Thus, we can interpret \code{stanhf} models as having implicit flat priors for parameters between any bounds placed on them in the \code{HistFactory} model, and constraint terms that are part of the likelihood.

If we desired other choices of prior, we could code them in the \code{Stan} language and rebuild the model. E.g., for a Gaussian prior for the parameter \code{mu}, $\mu \sim \mathcal{N}(1, 0.1)$, we could add
\begin{lstlisting}[language=Stan]
model {
    // ... other model statements omitted ...
    mu ~ normal(1., 0.1);
}
\end{lstlisting}
to the \code{model} block. These choices cannot be automated and should be considered by a user.

Lastly, we may wish to choose a conjugate prior~\cite{raiffa2000applied} --- that is, a prior $\pr(\param)$ such that the prior and posterior $\pr(\param \given K)$ belong to the same family of distributions. In this case, we may update using the auxiliary data $K$ as a preliminary step and specify only the posterior in the \code{Stan} model. That is, we could remove the constraint term $\pr(K \given \param)$ and replace it with a posterior $\pr(\param \given K)$.

\subsection{Command-line options}

The behavior of the \code{stanhf} program can be altered through command-line options. E.g., the compilation and validation steps may be disabled. To see these flags, use the \code{{-}{-}help} flag:
\begin{lstlisting}[style=bash]
|\terminal| stanhf --help
\end{lstlisting}
\smallskip
\begin{lstlisting}[style=bash]
Usage: stanhf [OPTIONS] HF_FILE_NAME

  Convert, build and validate a histfactory json file HF_FILE_NAME as a Stan
  model.

Options:
  --version                       Show the version and exit.
  --build / --no-build            Build Stan program.
  --validate-par-names / --no-validate-par-names
                                  Validate Stan program parameter names.
  --validate-target / --no-validate-target
                                  Validate Stan program target.
  --cmdstan-path                  Show path to cmdstan.
  --patch <path to patchset> <number>
                                  Apply a patch to the model.
  -h, --help                      Show this message and exit.

  Check out |\href{https://github.com/xhep-lab/stanhf}{\textcolor{black}{https://github.com/xhep-lab/stanhf}}| for more details or to report
  issues
\end{lstlisting}

\subsection{Patches}

Patches are a way to modify a \code{HistFactory} model by e.g.~adding a specific signal. They are supported in \code{stanhf} by an optional argument \code{patch}. E.g., we can patch the model \code{RegionA/Bkgonly.json} from \cref{sec:quickstart}
\begin{lstlisting}[style=bash]
|\terminal| stanhf RegionA/Bkgonly.json --patch RegionA/patchset.json 0
\end{lstlisting}
This applies the first patch in the \code{RegionA/patchset.json} collection of patches to \code{RegionA/Bkgonly.json} and converts the resulting model to \code{Stan}. We use \code{pyhf} to apply patches.

\subsection{Null parameters and modifiers}

To simplify models, \code{stanhf} ignores modifiers and parameters that cannot influence the expected numbers of events. This occurs when, e.g., a modifier varies between \code{lo} and \code{hi}, but \code{lo == hi}. These modifiers are reported as being null. Parameters that are associated only with null modifiers are themselves null.

\subsection{Validation and testing}

The \code{stanhf} interpretation of a \code{HistFactory} model is validated against \code{pyhf} by 
\begin{enumerate}[leftmargin=*]
    \item Checking that the parameter names and sizes agree; see \code{stanhf.Converter.validate\_par\_names}.
    \item Checking that the log-likelihoods agree up to a constant term.
    This is performed by checking that
        \begin{equation*}
        \Delta\log\mathcal{L}_\texttt{stanhf} = \Delta\log\mathcal{L}_\texttt{pyhf} 
    \end{equation*}
    The log-likelihoods are evaluated at randomly chosen points found by perturbing the suggested initial choices by noise. See \code{stanhf.Converter.validate\_target}.
\end{enumerate}
A constant term in the log-likelihood cannot impact statistical inferences currently available in \code{pyhf} or \code{Stan} and, for performance, is the default behavior in \code{Stan}.\footnote{Though this constant can be relevant in model selection if it differs between models.}

As part of the \code{stanhf} testing set, these checks are performed for several models and patches available from \code{HEPData}; see e.g.~\code{stanhf/tests/test\_hepdata.py}. These tests, and others, can be ran using 
\begin{lstlisting}[style=bash]
|\terminal| pytest .
\end{lstlisting}
from within the \code{stanhf} source code.

\section{Bayesian inference}\label{sec:bayes}

The \code{stanhf} CLI builds an executable \code{Stan} program and writes associated data files. We briefly describe how to use them for Bayesian inference. For concreteness, we look here at the \code{examples/normfactor.json} model shown in \cref{fig:example_normfactor}. In this model, a signal is characterized by a strength parameter $\mu$ and $\mu = 0$ corresponds to the background-only model. First, we must build the model and sample from it,
\begin{lstlisting}[language=bash]
|\terminal| stanhf examples/normfactor.json  # make stan program
|\terminal| examples/model sample num_chains=4 data file=examples/normfactor_data.json init=examples/normfactor_init.json  # run stan program
\end{lstlisting}
This uses the HMC sampling algorithm with 4 chains, in conjunction with the \code{stanhf} data files. This is the \code{cmdstan} interface to \code{Stan}, where we run \code{Stan} at the command-line; you can alternatively explore language-specific interfaces. There are \code{Python}, \code{Julia}, \code{R} and \code{MATLAB} interfaces~\cite{interfaces}, as well as a web interface~\cite{stanplayground}. For all CLI options, see \code{examples/model {-}{-}help}.

From the CLI interface, the output from \code{Stan} shows,
\begin{lstlisting}[style=bash]
sample
    num_samples = 1000 (Default)
|\termskip| 
    algorithm = hmc (Default)
|\termskip| 
data
  file = examples/normfactor_data.json
init = examples/normfactor_init.json
random
  seed = 3094244886 (Default)
output
  file = output.csv (Default)
|\termskip| 
Chain [4] Iteration:    1 / 2000 [  0%]  (Warmup)
|\termskip| 
Chain [4] Iteration: 1000 / 2000 [ 50%]  (Warmup)
Chain [4] Iteration: 1001 / 2000 [ 50%]  (Sampling)
|\termskip| 
Chain [4] Iteration: 2000 / 2000 [100%]  (Sampling)
|\termskip| 
\end{lstlisting}
\code{Stan} programs write Markov Chain Monte Carlo (MCMC) chains to plain text \code{csv} files, by default called \code{output\_\{chain number\}.csv}. 
There is a rich ecosystem of analysis and visualization software for MCMC chains, including \code{Stan} outputs, such as
\code{ArviZ}~\cite{arviz_2019},
\code{BayesPlot}~\cite{bayesplot,bayes-viz} and
\code{ShinyStan}~\cite{cmdstanr,shinystan}. We show examples of using them all in \code{stanhf/EXAMPLE.md}. 

Here we demonstrate \code{ArviZ}, since it is a \code{Python} package, and most familiar to users in high-energy physics. After installing \code{ArviZ},
\begin{lstlisting}[language=bash]
|\terminal| pip install arviz
\end{lstlisting}
you can analyze the \code{Stan} results. We can find the mean, standard deviation and credible region of the strength parameter,
\begin{lstlisting}[style=python]
|\repl| import arviz as az
|\repl| data = az.from_cmdstan('output_*.csv')
|\repl| az.summary(data, 'mu', kind='stats')
|\termout\verb|      mean     sd  hdi_3%  hdi_97%||
|\termout\verb| mu  0.516  0.415     0.0    1.272||
\end{lstlisting}
We may check the effective sample size and $\hat R$ diagnostic~\cite{Vehtari_2021},
\begin{lstlisting}[style=python]
|\repl| az.summary(data, 'mu', kind='diagnostics')
|\termout\verb|     mcse_mean  mcse_sd  ess_bulk  ess_tail  r_hat||
|\termout\verb| mu      0.012    0.008     773.0     682.0    1.0||
\end{lstlisting}
We can plot posterior distributions. For example, to show the one-dimensional posterior pdf
\smallskip
\begin{mdframed}[
    backgroundcolor=white,
    linewidth=1pt,%
    linecolor=gray,
    innerleftmargin=-1pt,
    innertopmargin=-3pt,
    innerbottommargin=-6pt,
    innerrightmargin=4pt]
\begin{lstlisting}[style=python,frame=none,aboveskip=1.25\baselineskip]
|\repl| import matplotlib.pyplot as plt
|\repl| az.plot_dist(data['posterior']['mu'])
|\repl| plt.ylabel('posterior pdf (a.u.)')
|\repl| plt.xlabel('mu')
|\repl| plt.show()
|
\vspace{-0.2cm}
\begin{center}
\includegraphics[width=0.9\linewidth]{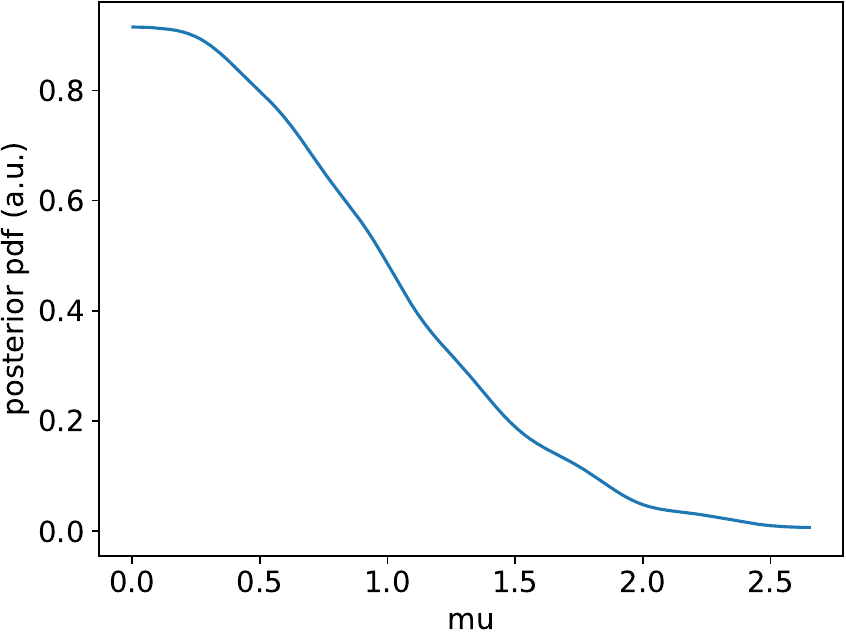}
\end{center}
\vspace{0.5cm}
|
\end{lstlisting}
\end{mdframed}
Lastly, the Bayes factor~\cite{Kass1995} and Bayes factor surface~\cite{Fowlie:2024dgj} can be computed using a Savage-Dickey ratio~\cite{Dickey1970}. The Bayes factor surface as a function of $\mu$ relative to the background-only model could be found by,
\smallskip
\begin{mdframed}[
    backgroundcolor=white,
    linewidth=1pt,%
    linecolor=gray,
    innerleftmargin=-1pt,
    innertopmargin=-3pt,
    innerbottommargin=-6pt,
    innerrightmargin=4pt]
\begin{lstlisting}[style=python,frame=none,aboveskip=1.25\baselineskip]
|\repl| import arviz as az
|\repl| from stanhf.contrib.bfs import plot_bfs

|\repl| data = az.from_cmdstan('output_*.csv')
|\repl| plot_bfs(data, 'mu', show=True)
|
\vspace{-0.2cm}
\begin{center}
\includegraphics[width=0.9\linewidth]{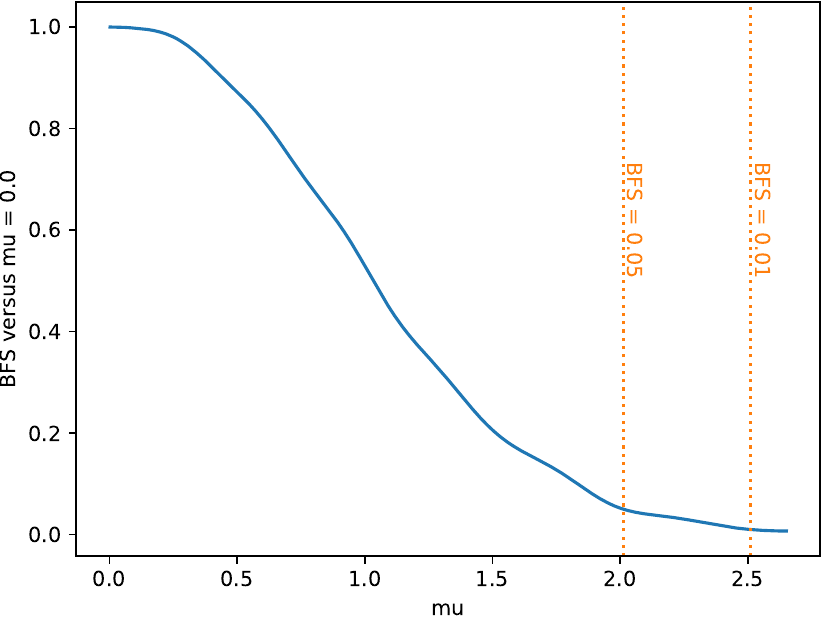}
\end{center}
\vspace{0.5cm}
|
\end{lstlisting}
\end{mdframed}
This calculation assumed that we sampled the posterior using a flat prior for $\mu$, though the result is independent of that choice.\footnote{Other priors are supported using the \code{prior} keyword argument.} To further explore Bayesian workflows, this model can be compiled, run and analyzed online \normfactorpg{here} using the \code{Stan} online playground~\cite{stanplayground}.

\section{Frequentist inference}\label{sec:freq}

As well as using \code{stanhf} for Bayesian inference using the powerful HMC functionality of \code{Stan}, we can harness the expressive power of \code{Stan} models in a frequentist setting. \code{stanhf} provides classes and functions to use a \code{Stan} model and optimizer in place of a \code{pyhf} model or native \code{pyhf} optimizer in the statistical inference package of \code{pyhf.infer}, which uses results in \refcite{Cowan:2010js}.\footnote{Frequentist inference is possible in \code{Stan} itself using simulation; see e.g.~\refcite{bootstrap}.}

As an example, we again use \code{examples/normfactor.json}, which declares that \code{mu} is the parameter of interest (POI). Using \code{pyhf}, we can plot the $\text{CL}_s$ statistic~\cite{Read2002} as a function of $\mu$:
\smallskip
\begin{mdframed}[
    backgroundcolor=white,
    linewidth=1pt,%
    linecolor=gray,
    innerleftmargin=-1pt,
    innertopmargin=-3pt,
    innerbottommargin=-6pt,
    innerrightmargin=4pt]
\begin{lstlisting}[style=python,frame=none, ,aboveskip=1.25\baselineskip]
|\repl| from pyhf.infer import hypotest
|\repl| from pyhf.contrib.viz import brazil

|\repl| from stanhf.contrib.freq import MockPyhfModel, mock_pyhf_backend
|\repl| from stanhf import Convert

|\repl| import numpy as np

|\repl| convert = Convert('examples/model.json')
|\repl| stan_file_name, data_file_name, init_file_name = convert.write_to_disk()

|\repl| model = MockPyhfModel(stan_file_name, data_file_name, init_file_name)

|\repl| test_poi = np.linspace(0., 5., 61)

|\repl| with mock_pyhf_backend():
|\context| results = [hypotest(t, model.config.data, model, return_expected_set=True) for t in test_poi]

|\repl| brazil.plot_results(test_poi, results)
|
\vspace{-0.2cm}
\begin{center}
\includegraphics[width=0.9\linewidth]{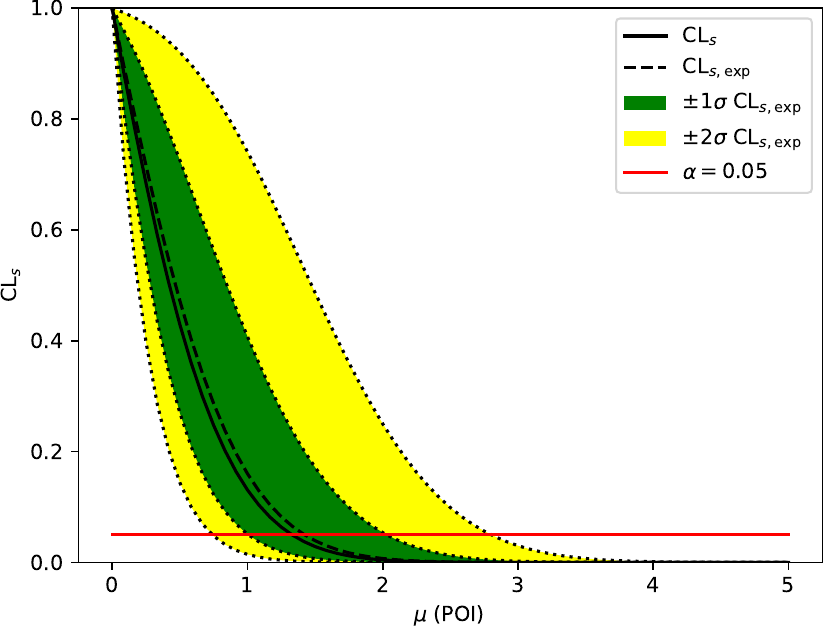}
\end{center}
\vspace{0.5cm}
|
\end{lstlisting}
\end{mdframed}
Here, \code{mock\_pyhf\_backend} provides a context in which \code{pyhf} uses the \code{Stan} optimizer as the backend, and \code{MockPyhfModel} creates a model that can be used in conjunction with the \code{Stan} optimizer. In this example, we see an upper limit of $1.3$ at $95\%$ for the parameter \code{mu}.

Thus, we can tweak models in \code{Stan} --- going beyond the restrictions of the declarative \code{HistFactory} specification --- and still access \code{pyhf} statistical machinery. In this case, however, it is a user's responsibility to ensure that any assumptions or approximations required by asymptotic formulae in \code{pyhf.infer} are valid.

The frequentist analysis in \code{pyhf.infer} requires one to select a particular POI. To perform tests on particular values of the POI, it must be possible to fix the POI. Thus, if a POI is specified in a \code{HistFactory} model, \code{stanhf} writes \code{Stan} code that allows it to be fixed. For a POI named \code{\{poi\}}, this is implemented using the pattern:
\begin{lstlisting}[language=Stan]
data {
    // 0 do not fix POI; 1 fix POI
    int<lower=0, upper=1> fix_{poi}; 
    // value of POI if it is fixed
    real fixed_{poi};  
}
parameters {
    // parameter for POI if it is not fixed
    array[1 - fix_{poi}] real free_{poi}; 
}
transformed parameters {
    // get value of POI from data if fixed or
    // parameter if not fixed
    real {poi} = fix_{poi} ? fixed_{poi} : free_{poi}[1];  
}
\end{lstlisting}

\section{Tweaking a model}\label{sec:tweak}

Although the \code{HistFactory} declarative specification describes many common models, we may wish to go beyond it. For example, suppose we wish to model contributions to a signal, $s_1$ and $s_2$, that depend on different powers of a modifier, e.g.,
\begin{equation}\label{eq:tweak}
    s = \mu s_1 + \mu^2 s_2 
\end{equation}
This form could occur because of interference effects or because the contributions occur at different orders in perturbation theory. 

We achieve this as follows. We begin from a model in which two signal contributions share a correlated normalization factor. This model can be found in the \code{stanhf} source code in \code{examples/tweak.json}. We generate a \code{Stan} model using \code{stanhf examples/tweak.json}. Because \code{Stan} is an imperative language, we can tweak the \code{Stan} code generated by \code{stanhf} as we wish. Thus, we edit the resulting \code{tweak.stan} model to match the required form \cref{eq:tweak}. Specifically, we change the \code{transformed parameters} block such that the second contribution to the signal depends on the square of the normalization factor,
\begin{lstlisting}[language=Stan]
  expected_channel_signal_1 *= mu;
  |\coderemove{expected\_channel\_signal\_2 *= mu;}|
  expected_channel_signal_2 *= |\codeadd{square(mu)}|;
\end{lstlisting}
We show this process in full in \cref{fig:example_tweak}.

\begin{figure*}
\begin{tcolorbox}[blankest, watermark text=\code{stanhf}\\\vspace{-4.5mm}\\$\scaleobj{1.5}{\Longrightarrow}$,
  watermark zoom=0.55,
  watermark color=violet,
  watermark opacity=0.25,
  clip watermark]
\begin{tcbraster}[raster columns=2,raster valign=top]
\begin{codebox}{HistFactory --- tweak.json \vphantom{Stan --- tweak.stan}}
\begin{lstlisting}[language=JSON,frame=none]
{
    "channels": [
        { "name": "channel",
          "samples": [
            { "name": "signal_1",
              "data": [50.0, 60.0],
              "modifiers": [
                { "name": "mu", "type": "normfactor", "data": null}
              ]
            },
            { "name": "signal_2",
              "data": [5.0, 10.0],
              "modifiers": [
                { "name": "mu", "type": "normfactor", "data": null}
              ]
            }
          ]
        }
    ],
    "observations": [
        { "name": "channel", "data": [44.0, 62.0] }
    ],
    "measurements": [
        { "name": "Measurement", "config":
          {"poi": "mu", "parameters": [ 
            { "name":"mu", "bounds": [[0.0, 10.0]], "inits": [1.0] }
          ]
          }
        }
    ],
    "version": "1.0.0"
}
\end{lstlisting}%
\end{codebox}%
\begin{tcolorbox}[blankest]
\begin{tcbraster}[raster columns=1]
\begin{codebox}{Stan --- tweak.stan \vphantom{HistFactory --- tweak.json}}%
\begin{lstlisting}[style=stanhf,frame=none]
data {
  vector[2] nominal_channel_signal_1;
  vector[2] nominal_channel_signal_2;
  int<lower=0, upper=1> fix_mu;
  real fixed_mu;
  tuple(real, real) lu_mu;
  array[2] int observed_channel;
}
parameters {
  array[1 - fix_mu] real<lower=lu_mu.1, upper=lu_mu.2> free_mu;
}
transformed parameters {
  vector[2] expected_channel_signal_1 = nominal_channel_signal_1;
  vector[2] expected_channel_signal_2 = nominal_channel_signal_2;
  real mu = fix_mu ? fixed_mu : free_mu[1];
  expected_channel_signal_1 *= mu;
  expected_channel_signal_2 *= mu;
  vector[2] expected_channel = expected_channel_signal_1 + expected_channel_signal_2; 
}
model {
  observed_channel ~ poisson(expected_channel);
}
generated quantities {
  array[2] int rv_expected_channel = poisson_rng(expected_channel);
}
\end{lstlisting}
\end{codebox}

\begin{codebox}{Edit \code{transformed parameters} block by hand \vphantom{HistFactory --- tweak.json}}%
\begin{lstlisting}[style=stanhf,frame=none]
transformed parameters {
  vector[2] expected_channel_signal_1 = nominal_channel_signal_1;
  vector[2] expected_channel_signal_2 = nominal_channel_signal_2;
  real mu = fix_mu ? fixed_mu : free_mu[1];
  expected_channel_signal_1 *= mu;
  |\coderemove{expected\_channel\_signal\_2 *= mu;}|
  expected_channel_signal_2 *= |\codeadd{square(mu)}|;
  vector[2] expected_channel = expected_channel_signal_1 + expected_channel_signal_2; 
}
\end{lstlisting}
\end{codebox}

\end{tcbraster}
\end{tcolorbox}
\end{tcbraster}
\end{tcolorbox}
\caption{Example of tweaking \code{stanhf} output to go beyond the \code{HistFactory} declarative specification. For simplicity, we omit the \code{functions} block of the \code{Stan} model and stripped comments and metadata from \code{stanhf} outputs. This model can be found in the \code{stanhf} source code in \code{examples/tweak.json}.}
\label{fig:example_tweak}
\end{figure*}

\section{Conclusions}

We introduced \code{stanhf} --- a new way to analyze and develop statistical models based on histogrammed data. The code transpiles \code{HistFactory} models into \code{Stan}, a probabilistic programming language. This allows state-of-the-art, scalable optimization and sampling using automatic differentiation and HMC. Because \code{Stan} is an imperative language, \code{HistFactory} modeling choices can be tweaked and adapted, as desired.

\code{Stan} is part of an ecosystem of Bayesian analysis software, allowing MCMC chains to be scrutinized for convergence and plotted, and even allowing models to be reran and analyzed online. Lastly, to ensure that the frequentist toolbox remains available, we provide an interface between \code{Stan} models and the inference machinery of \code{pyhf}.

\begin{acknowledgments}
AF was supported by RDF-22-02-079.
\end{acknowledgments}

\bibliographystyle{JHEP}
\bibliography{refs}

\end{document}